\documentclass[a4paper,10pt]{article}
\usepackage{amsfonts}
\usepackage{amsmath}
\usepackage{amssymb}
\usepackage{graphicx}

\begin{document}

\title{Dynamical diffusion and renormalization group equation for the Fermi
velocity in doped graphene}
\author{J. S. Ardenghi\thanks{%
email:\ jsardenghi@gmail.com, fax number:\ +54-291-4595142}, P. Bechthold$%
^{\dag }$, P. Jasen$^{\dag }$, E. Gonzalez$^{\dag }$ and A. Juan$^{\dag }$ \\
IFISUR, Departamento de F\'{\i}sica (UNS-CONICET)\\
Avenida Alem 1253, Bah\'{\i}a Blanca, Argentina}
\maketitle

\begin{abstract}
The aim of this work is to study the electron transport in graphene with
impurities by introducing a generalization of linear response theory for
linear dispersion relations and spinor wave functions. Current response and
density response functions are derived and computed in the Boltzmann limit
showing that in the former case a minimum conductivity appears in the
no-disorder limit. In turn, from the generalization of both functions, an
exact relation can be obtained that relates both. Combining this result with
the relation given by the continuity equation, is possible to obtain general
functional behavior of the the diffusion pole. Finally, a dynamical
diffusion is computed in the quasistatic limit using the definition of
relaxation function. A lower cutoff must be introduce to regularize infrared
divergences which allows to obtain a full renormalization group equation for
the Fermi velocity, which is solved up to order $O(\hbar ^{2})$.
\end{abstract}

\section{Introduction}

Graphene is a two dimensional hexagonal lattice of carbon atoms and is one
of the most important topics in solid state physics due to the vast
application in nano-electronics, opto-electronics, superconductivity and
Josephson junctions (\cite{novo},\cite{intro1},\cite{intro2},\cite{B} and 
\cite{BBBB}). The band structure shows that the conduction and valence band
touch at the Dirac point and the dispersion relation is approximately linear
and isotropic \cite{A}. This linear dispersion near the symmetry points have
striking similarities with those of massless relativistic Dirac fermions 
\cite{B}. This leads to a number of fascinating phenomena such as the
half-quantized Hall effect (\cite{C},\cite{D}) and minimum quantum
conductivity in the limit of vanishing concentration of charge carriers \cite%
{novo}. Although this is an outstanding experimental result, there is no
consensus about the theoretical value computed through different theoretical
methods (see \cite{Ziegler}), neither the physical reason for such minimum
value (see \cite{yuriv}), where the minimum is due to the impurity resonance
and is not related to the Dirac point.

In particular, one of the theoretical methods used to compute response
functions within the linear response theory is the Kubo formalism \cite{kubo}%
. Deviations of charge and current densities from their equilibrium values
are described by density and current response functions through Kubo
formulas using the same two-particle Green function. Although, generally is
unable to obtain exact relations between these response functions, several
approximations can be obtained taking into account the dimensionality and
dispersion relation of the system (see \cite{Janis}). But these approximate
relations are based on the continuity equation and Ward identities and is
not clear if these assumptions are valid for linear dispersion relations and
spinor wave functions.

In turn, impurities in graphene can be considered in various type of forms:
substitutional, where the site energy is different from those of carbon
atoms, which originates resonances \cite{mahan} and as adsorbates, that can
be placed on various points in graphene; sixfold hollow site of a honeycomb
lattice, twofold bridge site of two neighboring carbons or top site of a
carbon atom \cite{roten}. Theoretical as well as experimental studies have
indicated that substitutional doping of carbon materials can be used to
tailor their physical and/or chemical properties (\cite{strobel}, \cite%
{jiang}, \cite{sankaran}). In particular, nitrogen or boron dopants can be
added to pristine graphene (\cite{11},\cite{12},\cite{13},\cite{14}).

The detection and absorption of low levels of hydrogen becomes very
important for sensor gas and hydrogen energy. Different methods of hydrogen
detection are not entirely selective or it have a high cost of manufacture
due to their complexity. Pd-doped reduced graphene have a clear response to
hydrogen and are very selective (\cite{pandey}, \cite{corral}). In the other
side, the decoration of carbon support by transition metals can also be
independently used to enhance the hydrogen storage of the specimens.
Transition metals eliminate the hydrogen dissociation barrier altogether 
\cite{adams}.

In this sense, the density and current response function of doped-graphene
with low concentration of subtitutional impurities is of major importance
for the consequences in the sensor effect (\cite{sch}, \cite{bar}). In
particular, the simplest graphene-based sensor detects the conductivity
change upon adsorption of analyte molecules. The change of conductivity
could be attributed to the changes of charge carrier concentration in the
graphene induced by adsorbed gas molecules. It has been proposed that such
device may be capable of detecting individual molecule \cite{kong}. These
reactions release captured electrons in the interaction zone between the gas
and the sensor, and increase their concentration in the conductivity zone.
But the conductivity of electrons are based on the diffusion phenomena of
charge carriers through the sample. Electrons moving in randomly distributed
scatterers has a diffusive character, which is described at long distances
by a diffusion equation. It has been shown that it is possible to supress
diffusion (see \cite{Anderson}), giving rise to a localization phenomena,
which will affect the sensor characteristics of the material. In turn, a
dynamical generalization of the diffusion constant from the electron-hole
correlation function cannot be linked to the frequency dependent
conductivity (see eq.(3.18) and eq.(3.19) of \cite{Janis}). In this sense,
the aim of this work is two-fold: to introduce a generalization of the
linear response theory for linear dispersion relation and spinor wave
functions, to apply it to graphene, and the subsequent computation of
minimal conductivity and dynamical diffusion, to analize the general
behavior of the system under local perturbations and the implications for
sensor gas.

This work will be organized as follow: In section II, the impurity averaged
Green function will be computed. In section III and IV, a generalization of
the conductivity tensor and response function will be computed using the
current definition for relativistic Dirac fermions. In section V, different
limit behavior of the current and density response functions are computed.
The Boltzmann limit is introducedshowing the minimal conductivity value. In
section VI, the dynamical diffusion will be computed through the relaxation
function, showing how the obtain the full renormalization group equation for
the Fermi velocity. Finally, the conclusion are presented. Appendix A and B
are introduced for self-contained lecture.

\section{Impurity averaged Green function}

The Hamiltonian of clean graphene in the $K$ point in the Brillouin zone and
in the long wavelength approximation reads (see \cite{B})%
\begin{equation}
H=v_{F}\left( 
\begin{array}{cc}
0 & p_{x}+ip_{y} \\ 
p_{x}-ip_{y} & 0%
\end{array}%
\right)   \label{h1}
\end{equation}%
where $v_{F}\sim 10^{6}m/s$ is the Fermi velocity. The eigenfunctions of
this Hamiltonian reads%
\begin{equation}
\psi _{\mathbf{k}}(\mathbf{r})=\frac{1}{\sqrt{2}}\left( 
\begin{array}{c}
1 \\ 
\lambda e^{i\varphi _{\mathbf{k}}}%
\end{array}%
\right) e^{i\mathbf{k}\cdot \mathbf{r}}  \label{h2}
\end{equation}%
where%
\begin{equation}
\varphi _{\mathbf{k}}=\arctan (\frac{k_{y}}{k_{x}})  \label{h3}
\end{equation}%
In turn, the eigenvalues reads%
\begin{equation}
E_{\lambda }(\mathbf{k})=\lambda v_{f}\hbar k  \label{h4}
\end{equation}%
where $k=\left\vert \mathbf{k}\right\vert $ and where $\lambda =1$ are
positive energy states (conduction band) and $\lambda =-1$ are negative
energy states (valence band).With the eigenfunctions of eq.(\ref{h2}) we can
compute the retarded and advanced Green function for conduction electrons $%
(\lambda =1)$ in momentum space\footnote{%
We assume that the valence-band states do not contribute to low temperature
conductivity.} 
\begin{equation}
G_{0}^{R(A)}(\mathbf{q},E)=\frac{1}{E-v_{f}\hbar q\mp is}\left( 
\begin{array}{cc}
1 & e^{i\varphi _{\mathbf{q}}} \\ 
e^{-i\varphi _{\mathbf{q}}} & 1%
\end{array}%
\right)   \label{h7}
\end{equation}%
where the minus sign correspond to the retarded Green function and the plus
sign to the advanced Green function. The contribution to second order in the
perturbation expansion in the impurity potential reads (see \cite{Rammer},
eq.(3.31), page 136)%
\begin{equation}
\left\langle G_{2}^{R}(\mathbf{k},\mathbf{k}^{\prime },E)\right\rangle
^{i=j}=\delta _{\mathbf{k,k}^{\prime }}n_{i}[G_{0}^{R}(\mathbf{k}%
,E)]^{2}\int \frac{d^{2}\mathbf{k}^{\prime }}{(2\pi )^{2}}\left\vert V_{imp}(%
\mathbf{k}-\mathbf{k}^{\prime })\right\vert ^{2}IG_{0}^{R}(\mathbf{k}%
^{\prime },E)  \label{h8}
\end{equation}%
where $n_{i}$ is the impurity concentration, $\left\vert V_{imp}(\mathbf{k}-%
\mathbf{k}^{\prime })\right\vert ^{2}I$ is a diagonal matrix 
\begin{equation}
\left\vert V_{imp}(\mathbf{k}-\mathbf{k}^{\prime })\right\vert ^{2}I=\left( 
\begin{array}{cc}
\left\vert V_{imp}(\mathbf{k}-\mathbf{k}^{\prime })\right\vert ^{2} & 0 \\ 
0 & \left\vert V_{imp}(\mathbf{k}-\mathbf{k}^{\prime })\right\vert ^{2}%
\end{array}%
\right)   \label{h9}
\end{equation}%
and the angle brackets represent the configurational averaging that can be
computed as%
\begin{equation}
\left\langle A\right\rangle =\int \overset{N}{\underset{i=1}{\prod }}d%
\mathbf{r_{i}}A(\mathbf{r_{1},r_{2},...,r_{N}})P(\mathbf{%
r_{1},r_{2},...,r_{N}})  \label{h9.1}
\end{equation}%
where $P(\mathbf{r_{1},r_{2},...,r_{N}})=P(\mathbf{r_{1}})P(\mathbf{r_{2}}%
)...P(\mathbf{r_{N}})$ and $P(\mathbf{r_{i}})$ is the probability density
for having the impurity located around point $\mathbf{r_{i}}$.\footnote{%
In this case we are assuming that the positions of the impurities are
distributed independently.}\ In eq.(\ref{h8}), the Fourier transform of $%
G_{2}^{R}(\mathbf{r},\mathbf{r}^{\prime },E)~$has been taken first.
Replacing last equation and eq.(\ref{h7}) in eq.(\ref{h8}) the diagonal part
of the averaged Green function reads%
\begin{gather}
\int \frac{d^{2}\mathbf{k}^{\prime }}{(2\pi )^{2}}\frac{\left\vert V_{imp}(%
\mathbf{k}-\mathbf{k}^{\prime })\right\vert ^{2}}{E-v_{f}\hbar k^{\prime }-is%
}=  \label{h11} \\
\int \frac{d^{2}\mathbf{k}^{\prime }}{(2\pi )^{2}}\left\vert V_{imp}(\mathbf{%
k}-\mathbf{k}^{\prime })\right\vert ^{2}\left( \frac{E-v_{f}\hbar k^{\prime }%
}{(E-v_{f}\hbar k^{\prime })^{2}+s^{2}}+i\frac{s}{(E-v_{f}\hbar k^{\prime
})^{2}+s^{2}}\right)   \notag
\end{gather}%
If we consider for simplicity that the impurity potential is a Dirac delta
potential, then\footnote{%
In this case, the disorder introduced by the delta Dirac impurity potential
is an on-site diagonal disorder.}%
\begin{equation}
V_{imp}(\mathbf{k})=\int d^{2}\mathbf{r}e^{-i\mathbf{r}\cdot \mathbf{q}%
}V_{imp}(\mathbf{r})=\int d^{2}\mathbf{r}e^{-i\mathbf{r}\cdot \mathbf{q}%
}V_{0}\delta (\mathbf{r})=V_{0}  \label{h11.1}
\end{equation}%
Using the last result, the integral of eq.(\ref{h11}) reads%
\begin{gather}
V_{0}^{2}\underset{s\rightarrow 0}{\lim }\int \frac{d^{2}\mathbf{k}^{\prime }%
}{(2\pi )^{2}}\left( \frac{E-v_{f}\hbar k^{\prime }}{(E-v_{f}\hbar k^{\prime
})^{2}+s^{2}}+i\frac{s}{(E-v_{f}\hbar k^{\prime })^{2}+s^{2}}\right) =
\label{h13} \\
i\pi V_{0}^{2}\int \frac{d^{2}\mathbf{k}^{\prime }}{(2\pi )^{2}}\delta
(E-v_{f}\hbar k^{\prime })=i\pi V_{0}^{2}n(E_{F})  \notag
\end{gather}%
where we have used that $\delta (x)=\frac{1}{\pi }\underset{s\rightarrow 0}{%
\lim }\frac{s}{x^{2}+s^{2}}$ and $n(E)$ is the density of states at the
Fermi energy.\footnote{%
In last equation the real part of is strictly not zero, but is a constant
that do not depends on the momentum. In this sense, this value is arbitrary
and has no observable consequences. For this we can assume that is zero or
redefine the reference for measuring energy.} At this point is important to
notice that in clean graphene, the density of states $n(E)$ at the Fermi
energy is $n(E_{F})=0$ (see \cite{Peres}, eq.(33)). Nevertheless, when
impurities are introduced, the density of states at the Fermi energy is not
zero (see \cite{Peres}, figure 3), which implies that disorder introduce an
imaginary term to the self-energy.

The non-diagonal term reads%
\begin{gather}
\int \frac{d^{2}\mathbf{k}^{\prime }}{(2\pi )^{2}}\frac{V_{0}^{2}}{%
E-v_{f}\hbar k^{\prime }-is}\frac{k_{x}^{\prime }+ik_{y}^{\prime }}{%
k^{\prime }}  \label{h14} \\
\int \frac{d^{2}\mathbf{k}^{\prime }}{(2\pi )^{2}}\frac{V_{0}^{2}}{k^{\prime
}}\left( \frac{k_{x}^{\prime }(E-v_{f}\hbar k^{\prime })-k_{y}^{\prime }s}{%
(E-v_{f}\hbar k^{\prime })^{2}+s^{2}}+i\frac{sk_{x}^{\prime }+k_{y}^{\prime
}(E-v_{f}\hbar k^{\prime })}{(E-v_{f}\hbar k^{\prime })^{2}+s^{2}}\right)  
\notag
\end{gather}%
Introducing polar coordinates in the wave vector $\mathbf{k}^{\prime }$, $%
k_{x}^{\prime }=k^{\prime }\cos \lambda $ and $k_{y}^{\prime }=k^{\prime
}\sin \lambda $, is not difficult to show the last integral is zero due to
the cosine and sine functions, which are integrated between $0$ and $2\pi $.
Then, the averaged Green function at second order in the perturbation
expansion in the impurity potential reads 
\begin{equation}
\left\langle G_{2}^{R}(\mathbf{k},\mathbf{k}^{\prime },E)\right\rangle
^{i=j}=\delta _{\mathbf{k,k}^{\prime }}[G_{0}^{R}(\mathbf{k},E)]^{2}\left( 
\begin{array}{cc}
i\eta  & 0 \\ 
0 & i\eta 
\end{array}%
\right)   \label{h16}
\end{equation}%
where $\eta =\pi n_{i}V_{0}^{2}n(E)$. By introducing the one-particle
irreducible propagator, which correspond to all the diagrams which cannot be
cut in two by cutting an internal line, the impurity averaged propagator can
be written as a geometric series in terms of the self-energy (see \cite%
{Rammer}, page 141)%
\begin{equation}
G=G_{0}+G_{0}\Sigma G_{0}+...=G_{0}\sum\limits_{n=0}^{+\infty }(\Sigma
G_{0})^{n}=G_{0}\left( I-\Sigma G_{0}\right) ^{-1}  \label{h16.2}
\end{equation}%
where $\Sigma =\Sigma _{1}^{R}+\Sigma _{2}^{R}+...$ contains the
contributions at different orders in the perturbation expansion of the
impurity concentration. With the computation done in eq.(\ref{h16}) we
finally obtain%
\begin{equation}
G^{R(A)}(\mathbf{q})=G_{0}\left( I-\Sigma G_{0}\right) ^{-1}=\frac{1}{%
E-v_{f}\hbar q\mp is-i\eta }\left( 
\begin{array}{cc}
1 & e^{i\varphi _{\mathbf{q}}} \\ 
e^{-i\varphi _{\mathbf{q}}} & 1%
\end{array}%
\right)   \label{h16.5}
\end{equation}%
This last result is the impurity averaged Green function which take into
account the first contribution of the self-energy by comparing last equation
with eq.(\ref{h7}). This is known as the full Born approximation, which
include electronic scattering from a single impurity. The diagonal part
contains the shifted pole due to the imaginary part of the self energy. The
non-diagonal part contains the same contribution multiplied by a phase
factor. The last result will be used in the following sections.

\section{Current response function}

In this section, a generalization of the conducitivity tensor for Dirac
fermion systems, that is, linear dispersion relation and spinors wave
functions, will be introduced. To do it we will follow the development
introduced in \cite{Rammer} and by taking into account the differences
introduced by Dirac systems. The Hamiltonian of Bloch electrons in the long
wavelength approximation in a electric field and random impurities reads%
\begin{equation}
H=v_{f}\mathbf{\sigma }\cdot (\mathbf{p}-e\mathbf{A})+V_{imp}(\mathbf{r})
\label{con1}
\end{equation}%
where $\mathbf{A}(\mathbf{r})$ is the vector potential that is related to
the electric field as%
\begin{equation}
\mathbf{E}=-\frac{\partial \mathbf{A}}{\partial t}  \label{con2}
\end{equation}%
and where $V_{imp}(\mathbf{r})$ is the impurity field. We can compute the
current density to linear order in the external electric field (see
eq.(7.84) of \cite{Rammer}) 
\begin{equation}
\mathbf{j}(\mathbf{r},t)=Tr(\rho _{0}(t)\mathbf{j})-\frac{i}{\hbar }%
\int_{t_{i}}^{t}d\overline{t}Tr(\rho _{0}(t_{r})[\mathbf{j}_{p}(\mathbf{r}%
,t),H_{A}(\overline{t})]+O(E^{2})  \label{con9}
\end{equation}%
where the charge current density operator can be written as%
\begin{equation}
\mathbf{j}_{p}=v_{F}\left\vert \mathbf{r}\right\rangle \mathbf{\sigma }%
\left\langle \mathbf{r}\right\vert  \label{con5.1}
\end{equation}%
which is the usual definition of current in relativistic Dirac system, where 
$v_{F}$ plays the role of velocity of light and 
\begin{equation}
H_{A}(t)=ev_{f}\mathbf{\sigma }\cdot \mathbf{A}  \label{con5.2}
\end{equation}%
Taking into account the direction of the current in index notation and to
linear order in the electric field we obtain%
\begin{equation}
j_{\alpha }(\mathbf{r},t)=\left\langle j_{\alpha }(\mathbf{r}%
,t)\right\rangle _{0}+\sum\limits_{\beta }^{{}}\int d\mathbf{r}^{\prime
}\int_{t}^{+\infty }Q_{\alpha \beta }(\mathbf{r},t;\mathbf{r}^{\prime
},t^{\prime })A_{\beta }(\mathbf{r}^{\prime },t^{\prime })  \label{con10}
\end{equation}%
where $Q_{\alpha \beta }$ is the current response function. Taking into
accout that in linear response, each frequency contributes additively, only
is necesary to study what happens at one driving frequency%
\begin{equation}
\mathbf{A}(\mathbf{r},t)=\mathbf{A}(\mathbf{r},\omega )e^{-i\omega t}
\label{con13}
\end{equation}%
Then, the Fourier transform of the current reads%
\begin{equation}
j_{\alpha }(\mathbf{r},\omega )=\left\langle j_{\alpha }(\mathbf{r},\omega
)\right\rangle _{0}+\sum\limits_{\beta }^{{}}\int d\mathbf{r}^{\prime
}Q_{\alpha \beta }(\mathbf{r},\mathbf{r}^{\prime },\omega )A_{\beta }(%
\mathbf{r}^{\prime },\omega )  \label{con14}
\end{equation}%
where%
\begin{equation}
Q_{\alpha \beta }(\mathbf{r},\mathbf{r}^{\prime },\omega )=K_{\alpha \beta }(%
\mathbf{r},\mathbf{r}^{\prime },\omega )-K_{\alpha \beta }(\mathbf{r},%
\mathbf{r}^{\prime },0)  \label{con15}
\end{equation}%
and%
\begin{equation}
K_{\alpha \beta }(\mathbf{r},\mathbf{r}^{\prime },\omega
)=\sum\limits_{\lambda \lambda ^{\prime }}^{{}}\frac{\rho _{\lambda }-\rho
_{\lambda ^{\prime }}}{\epsilon _{\lambda }-\epsilon _{\lambda ^{\prime
}}+\hbar \omega +is}\left\langle \lambda \left\vert j_{\alpha }^{p}(\mathbf{r%
})\right\vert \lambda ^{\prime }\right\rangle \left\langle \lambda ^{\prime
}\left\vert j_{\beta }^{p}(\mathbf{r}^{\prime })\right\vert \lambda
\right\rangle  \label{con16}
\end{equation}%
where $\left\vert \lambda \right\rangle $ are eigenstates of unperturbed
Hamiltonian and $\rho _{\lambda }$ is the mean ocuppation number for a
energy level $\epsilon _{\lambda }$. At this point, if we use the usual
definition of current in non-relativistic quantum mechanics%
\begin{equation}
\mathbf{j}=\frac{e}{2m}\{\mathbf{P},\left\vert \mathbf{r}\right\rangle
\left\langle \mathbf{r}\right\vert \}  \label{con17}
\end{equation}%
then 
\begin{equation}
\left\langle \lambda \left\vert j_{\alpha }^{p}(\mathbf{r})\right\vert
\lambda ^{\prime }\right\rangle =-\frac{i\hbar e}{2m}[\psi _{\lambda
^{\prime }}^{\ast }(\mathbf{r})\overrightarrow{\nabla }\psi _{\lambda }(%
\mathbf{r})-\psi _{\lambda }(\mathbf{r})\overrightarrow{\nabla }\psi
_{\lambda ^{\prime }}^{\ast }(\mathbf{r})]  \label{con18}
\end{equation}%
In the same line of thought, we can use the definition of relativistic Dirac
current, then%
\begin{equation}
v_{F}\left\langle \lambda \mid \mathbf{r}\right\rangle \mathbf{\sigma }%
\left\langle \mathbf{r}\mid \lambda ^{\prime }\right\rangle =v_{F}\psi
_{\lambda }^{\dag }(\mathbf{r})\mathbf{\sigma }\psi _{\lambda ^{\prime }}(%
\mathbf{r})  \label{con19}
\end{equation}%
and in the same way%
\begin{equation}
\left\langle \lambda ^{\prime }\left\vert j_{\beta }^{p}(\mathbf{r}%
)\right\vert \lambda \right\rangle =v_{F}\psi _{\lambda ^{\prime }}^{\dag }(%
\mathbf{r}^{\prime })\mathbf{\sigma }\psi _{\lambda }(\mathbf{r}^{\prime })
\label{con20}
\end{equation}%
Introducing eq.(\ref{con19}) and eq.(\ref{con20}) into eq.(\ref{con16}) and
writing in index notation which allows to move the functions $\psi $ and the
Pauli matrices we have%
\begin{gather}
K_{\alpha \beta }(\mathbf{r},\mathbf{r}^{\prime },\omega
)=e^{2}v_{F}^{2}\int \frac{dE}{2\pi }\int \frac{dE^{\prime }}{2\pi }\frac{%
\rho (E)-\rho (E^{\prime })}{E-E^{\prime }+\hbar \omega +is}\psi
_{l}^{\lambda }(\mathbf{r}^{\prime })\psi _{i}^{\lambda \ast }(\mathbf{r}%
)\sigma _{ij}^{\alpha }\sigma _{kl}^{\beta }\psi _{j}^{\lambda ^{\prime }}(%
\mathbf{r})\psi _{k}^{\lambda ^{\prime }\ast }(\mathbf{r}^{\prime })=
\label{con20.2} \\
e^{2}v_{F}^{2}\int \frac{dE}{2\pi }\int \frac{dE^{\prime }}{2\pi }\frac{\rho
(E)-\rho (E^{\prime })}{E-E^{\prime }+\hbar \omega +is}A_{li}(\mathbf{r}%
^{\prime },\mathbf{r},E)\sigma _{ij}^{\alpha }\sigma _{kl}^{\beta }A_{jk}(%
\mathbf{r},\mathbf{r}^{\prime },E^{\prime })  \notag
\end{gather}%
where we have used the relation between the spectral weigth $A(\mathbf{r},%
\mathbf{r}^{\prime },E)$ and the wave functions (see Appendix A, eq.(\ref%
{con21}) and eq.(\ref{con22})). Finally, applying the relation between the
spectral weight and the Green function we obtain%
\begin{gather}
K_{\alpha \beta }(\mathbf{r},\mathbf{r}^{\prime },\omega
)=-e^{2}v_{F}^{2}\int \frac{dE}{2\pi }\int \frac{dE^{\prime }}{2\pi }\frac{%
\rho (E)-\rho (E^{\prime })}{E^{\prime }-E+\hbar \omega +is}\times
\label{con20.3} \\
\left[ G_{li}^{R}(\mathbf{r}^{\prime },\mathbf{r},E)-G_{li}^{A}(\mathbf{r}%
^{\prime },\mathbf{r},E)\right] \sigma _{ij}^{\alpha }\sigma _{kl}^{\beta }%
\left[ G_{jk}^{R}(\mathbf{r},\mathbf{r}^{\prime },E^{\prime })-G_{jk}^{A}(%
\mathbf{r},\mathbf{r}^{\prime },E^{\prime })\right]  \notag
\end{gather}%
which is the desired generalization of the current response function for
Dirac fermion systems. In this case, the Pauli matrices play the rol of
momentum in eq.(7.96) of \cite{Rammer}. In the momentum space, the
current-current response function reads%
\begin{gather}
K_{\alpha \beta }(\mathbf{q},\mathbf{q}^{\prime },\omega
)=e^{2}v_{F}^{2}\int \frac{d^{2}\mathbf{k}}{(2\pi )^{2}}\int \frac{d^{2}%
\mathbf{k}^{\prime }}{(2\pi )^{2}}\int \frac{dE}{2\pi }\int \frac{dE^{\prime
}}{2\pi }\frac{\rho (E)-\rho (E^{\prime })}{E^{\prime }-E+\hbar \omega +is}%
\times  \label{con20.6.1.0} \\
Tr(\left[ G^{R}(\mathbf{k}+\mathbf{q}^{\prime },\mathbf{k}^{\prime }+\mathbf{%
q},E)-G^{A}(\mathbf{k}+\mathbf{q}^{\prime },\mathbf{k}^{\prime }+\mathbf{q}%
,E)\right] \sigma ^{\alpha }\left[ G^{R}(\mathbf{k}^{\prime }-\mathbf{q},%
\mathbf{k}-\mathbf{q}^{\prime },E^{\prime })-G^{A}(\mathbf{k}^{\prime }-%
\mathbf{q},\mathbf{k}-\mathbf{q}^{\prime },E^{\prime })\right] \sigma
^{\beta })  \notag
\end{gather}%
Introducing the impurity averaging of two Green function (see \cite{Rammer},
eq.(8.3))%
\begin{gather}
\left\langle G^{R(A)}(\mathbf{k}+\mathbf{q}^{\prime },\mathbf{k}^{\prime }+%
\mathbf{q},E)G^{A(R)}(\mathbf{k}^{\prime }-\mathbf{q},\mathbf{k}-\mathbf{q}%
^{\prime },E^{\prime })\right\rangle =  \label{con20.6.1.2} \\
\delta _{\mathbf{q},\mathbf{q}^{\prime }}\left\langle G^{R(A)}(\mathbf{k}+%
\mathbf{q},\mathbf{k}^{\prime }+\mathbf{q},E)G^{A(R)}(\mathbf{k}^{\prime }-%
\mathbf{q},\mathbf{k}-\mathbf{q},E^{\prime })\right\rangle  \notag
\end{gather}%
computing one of the energy integration and exploting the analytical
properties of the averaged Green function we obtain (see \cite{Rammer}, page
283)%
\begin{equation}
K_{\alpha \beta }(\mathbf{q},\omega )=K_{\alpha \beta }^{AR}(\mathbf{q}%
,\omega )+K_{\alpha \beta }^{AA}(\mathbf{q},\omega )+K_{\alpha \beta }^{RR}(%
\mathbf{q},\omega )  \label{con20.6.1.3}
\end{equation}%
where%
\begin{equation}
K_{\alpha \beta }^{RA}(\mathbf{q},\omega )=e^{2}v_{F}^{2}\sigma
_{ij}^{\alpha }\sigma _{kl}^{\beta }\int \frac{dE}{2\pi i}[\rho (E+\hbar
\omega )-\rho (E)]\Phi _{lijk}^{AR}(E+\hbar \omega ,E,\mathbf{q})
\label{con20.6.1.4}
\end{equation}%
and%
\begin{equation}
K_{\alpha \beta }^{AA}(\mathbf{q},\omega )=-e^{2}v_{F}^{2}\sigma
_{ij}^{\alpha }\sigma _{kl}^{\beta }\int \frac{dE}{2\pi i}\rho (E+\hbar
\omega )\Phi _{lijk}^{AA}(E+\hbar \omega ,E,\mathbf{q})  \label{con20.6.1.5}
\end{equation}%
\begin{equation}
K_{\alpha \beta }^{RR}(\mathbf{q},\omega )=e^{2}v_{F}^{2}\sigma
_{ij}^{\alpha }\sigma _{kl}^{\beta }\int \frac{dE}{2\pi i}\rho (E)\Phi
_{lijk}^{RR}(E+\hbar \omega ,E,\mathbf{q})  \label{con20.6.1.6}
\end{equation}%
where the electron (hole)-electron (hole) correlation function $\Phi
_{lijk}^{ab}$ reads%
\begin{equation}
\Phi _{lijk}^{ab}(E+\hbar \omega ,E,\mathbf{q})=\int \frac{d^{2}\mathbf{k}}{%
(2\pi )^{2}}\int \frac{d^{2}\mathbf{k}^{\prime }}{(2\pi )^{2}}\left\langle
G_{li}^{a}(\mathbf{k}+\mathbf{q},\mathbf{k}^{\prime }+\mathbf{q},E+\hbar
\omega )G_{jk}^{b}(\mathbf{k}^{\prime }-\mathbf{q},\mathbf{k}-\mathbf{q}%
,E)\right\rangle  \label{con20.6.1.7}
\end{equation}%
The final conductivity tensor can be written in terms of the current
response function $K_{\alpha \beta }(\mathbf{q},\omega )$ (see \cite{Rammer}%
, eq.(8.51)) using the Kramer-Kronig relation%
\begin{equation}
\sigma _{\alpha \beta }(\mathbf{q},\omega )=\frac{K_{\alpha \beta }(\mathbf{q%
},\omega )-K_{\alpha \beta }(\mathbf{q},0)}{i\omega }  \label{con20.8}
\end{equation}%
As we can see in eq.(\ref{con20.6.1.0}), we have the multiplication of one
Green matrix functions with the Pauli matrix in the $\alpha $ direction and
the other with the Pauli matrix in the $\beta $ direction%
\begin{equation}
G_{li}^{R(A)}(\mathbf{k}+\mathbf{q},\mathbf{k}^{\prime }+\mathbf{q},E+\hbar
\omega )\sigma _{ij}^{\alpha }=M_{lj}^{R(A)}(\mathbf{k}+\mathbf{q},\mathbf{k}%
^{\prime }+\mathbf{q},E+\hbar \omega )  \label{con20.9}
\end{equation}%
\begin{equation}
G_{jk}^{R(A)}(\mathbf{k}^{\prime }-\mathbf{q},\mathbf{k}-\mathbf{q},E)\sigma
_{kl}^{\beta }=M_{jl}^{R(A)}(\mathbf{k}^{\prime }-\mathbf{q},\mathbf{k}-%
\mathbf{q},E)  \label{con20.10}
\end{equation}%
The two possible Pauli matrices are $\sigma ^{x}$ and $\sigma ^{y}$ and in
particular if we choose the direction of the Pauli matrix in such a way that 
$\sigma ^{\alpha }$ is $\sigma ^{\alpha }=\sigma ^{x}\cos \phi e_{x}+\sigma
^{y}\sin \phi e_{y}$ and $\sigma ^{\beta }=\sigma ^{x}\cos \theta
e_{x}+\sigma ^{y}\sin \theta e_{y}$, where $\phi $ and $\theta $ are angles
in real space, then%
\begin{equation}
\sigma ^{\alpha }=\left( 
\begin{array}{cc}
0 & e^{-i\phi } \\ 
e^{i\phi } & 0%
\end{array}%
\right) \text{ \ \ \ \ \ \ \ }\sigma ^{\beta }=\left( 
\begin{array}{cc}
0 & e^{-i\theta } \\ 
e^{i\theta } & 0%
\end{array}%
\right)  \label{con20.13}
\end{equation}%
At this point, we have to used the perturbation expansion of the product of
two matrix Green functions in the impurity concentration which has been
computed in last section.

\section{Density response function}

In a similar way, we can generalize the density response function $\chi (%
\mathbf{r},\mathbf{r}^{\prime },\omega )$ for linear dispersion and spinor
wave functions, which is defined as (see eq.(7.23) of \cite{Rammer})%
\begin{equation}
\chi (\mathbf{r},\mathbf{r}^{\prime },\omega )=-\sum\limits_{\lambda \lambda
^{\prime }}^{{}}\frac{\rho _{\lambda }-\rho _{\lambda ^{\prime }}}{\epsilon
_{\lambda }-\epsilon _{\lambda ^{\prime }}+\hbar \omega +is}\left\langle
\lambda \mid \mathbf{r}\right\rangle \left\langle \mathbf{r}\mid \lambda
^{\prime }\right\rangle \left\langle \lambda ^{\prime }\mid \mathbf{r}%
^{\prime }\right\rangle \left\langle \mathbf{r}^{\prime }\mid \lambda
\right\rangle  \label{dif1}
\end{equation}%
Using eq.(\ref{con22}) and computing the Fourier transform we obtain for the
density response function%
\begin{gather}
\chi (\mathbf{q},\mathbf{q}^{\prime },\omega )=\int \frac{d^{2}\mathbf{k}}{%
(2\pi )^{2}}\int \frac{d^{2}\mathbf{k}^{\prime }}{(2\pi )^{2}}\int \frac{dE}{%
2\pi }\int \frac{dE^{\prime }}{2\pi }\frac{\rho (E)-\rho (E^{\prime })}{%
E^{\prime }-E+\hbar \omega +is}\times  \label{dif4} \\
\left[ G_{ij}^{R}(\mathbf{k}+\mathbf{q}^{\prime },\mathbf{k}^{\prime }+%
\mathbf{q},E)-G_{ij}^{A}(\mathbf{k}+\mathbf{q}^{\prime },\mathbf{k}^{\prime
}+\mathbf{q},E)\right] \left[ G_{ji}^{R}(\mathbf{k}^{\prime }-\mathbf{q},%
\mathbf{k}-\mathbf{q}^{\prime },E^{\prime })-G_{ji}^{A}(\mathbf{k}^{\prime }-%
\mathbf{q},\mathbf{k}-\mathbf{q}^{\prime },E^{\prime })\right]  \notag
\end{gather}%
Introducing the impurity averaging of two Green function 
\begin{gather}
\chi (\mathbf{q},\omega )=\int \frac{d^{2}\mathbf{k}}{(2\pi )^{2}}\int \frac{%
d^{2}\mathbf{k}^{\prime }}{(2\pi )^{2}}\int \frac{dE}{2\pi }\int \frac{%
dE^{\prime }}{2\pi }\frac{\rho (E)-\rho (E^{\prime })}{E^{\prime }-E+\hbar
\omega +is}\times  \label{dif4.0.1} \\
Tr(A(\mathbf{k}+\mathbf{q},\mathbf{k}^{\prime }+\mathbf{q},E)A(\mathbf{k}%
^{\prime }-\mathbf{q},\mathbf{k}-\mathbf{q},E^{\prime }))  \notag
\end{gather}%
and computing one of the energy integration by exploting the analytical
properties of the averaged Green function we obtain%
\begin{equation}
\chi (\mathbf{q},\omega )=\chi ^{AR}(\mathbf{q},\omega )+\chi ^{AA}(\mathbf{q%
},\omega )+\chi ^{RR}(\mathbf{q},\omega )  \label{dif4.1}
\end{equation}%
where%
\begin{equation}
\chi ^{RA}(\mathbf{q},\omega )=\int \frac{dE}{2\pi i}\left[ \rho (E+\hbar
\omega )-\rho (E)\right] \Phi _{lijk}^{AR}(E+\hbar \omega ,E,q)
\label{dif4.2}
\end{equation}%
and%
\begin{equation}
\chi ^{AA}(\mathbf{q},\omega )=-\int \frac{dE}{2\pi i}\rho (E+\hbar \omega
)\Phi _{lijk}^{AA}(E+\hbar \omega ,E,\mathbf{q})  \label{dif4.3}
\end{equation}%
\begin{equation}
\chi ^{RR}(\mathbf{q},\omega )=\int \frac{dE}{2\pi i}\rho (E)\Phi
_{lijk}^{RR}(E+\hbar \omega ,E,\mathbf{q})  \label{dif4.4}
\end{equation}%
With this result, the generalization of the density response function for
linear dispersion relation and spinor wave function is obtained.

\section{Current and density response relations}

With the generalization of the current and density response functions for
linear dispersion relations and spinors, we can proceed to obtain a relation
between those functions. Introducing a Kronecker delta product $\delta
_{ij}\delta _{kl}$ in $\chi (\mathbf{q},\omega )$ and using eq.(\ref{con20.8}%
) we can write%
\begin{equation}
i\omega \sigma _{\alpha \beta }(\mathbf{q},\omega )+e^{2}v_{F}^{2}\chi (%
\mathbf{q},\omega )=\xi (\mathbf{q},\omega )-K_{\alpha \beta }(\mathbf{q},0)
\label{cu2.1}
\end{equation}%
where%
\begin{equation}
\xi (\mathbf{q},\omega )=Tr(\Lambda \Gamma )=\Lambda _{ijkl}^{\alpha \beta
}\Gamma _{ijkl}(\mathbf{q},\omega )  \label{cu4}
\end{equation}%
where%
\begin{equation}
\Lambda _{ijkl}^{\alpha \beta }=e^{2}v_{F}^{2}(\sigma _{ij}^{\alpha }\sigma
_{kl}^{\beta }+\delta _{ij}\delta _{kl})  \label{cu5}
\end{equation}%
and%
\begin{gather}
\Gamma _{ijkl}(\mathbf{q},\omega )=\int \frac{d^{2}\mathbf{k}}{(2\pi )^{2}}%
\int \frac{d^{2}\mathbf{k}^{\prime }}{(2\pi )^{2}}\int \frac{dE}{2\pi }\int 
\frac{dE^{\prime }}{2\pi }\frac{\rho (E)-\rho (E^{\prime })}{E^{\prime
}-E+\hbar \omega +is}\times  \label{cu6} \\
\left\langle \left[ G_{li}^{R}(\mathbf{k}+\mathbf{q},\mathbf{k}^{\prime }+%
\mathbf{q},E)-G_{li}^{A}(\mathbf{k}+\mathbf{q},\mathbf{k}^{\prime }+\mathbf{q%
},E)\right] \left[ G_{jk}^{R}(\mathbf{k}^{\prime }-\mathbf{q},\mathbf{k}-%
\mathbf{q},E^{\prime })-G_{jk}^{A}(\mathbf{k}^{\prime }-\mathbf{q},\mathbf{k}%
-\mathbf{q},E^{\prime })\right] \right\rangle  \notag
\end{gather}%
Relation eq.(\ref{cu2.1}) is analogous to the relation introduced in \cite%
{Janis}, eq.(38), but in the former case, the relation obtained is not a
definition as it occurs in \cite{Janis}. The main difference is that in
graphene and in general for spinor systems with linear dispersion relation,
the space derivate is replaced by the Pauli matrix, then the Fourier
transform do not introduce any momentum $\mathbf{p}$. The non-appearance of
the momentum in the current response function implies a different functional
behavior, but the same diagrammatic expansion. In the other side, whenever
there is a continuity equation, which expresses the charge conservation, it
is possible to obtain a direct relation between isotropic conductivity and
density response function 
\begin{equation}
\sigma (\mathbf{q},\omega )=-\frac{ie^{2}\omega }{q^{2}}\chi (\mathbf{q}%
,\omega )  \label{cu7}
\end{equation}%
In the particular case of graphene, a continuity equation can be obtained,
which is identical to continuity equation for quantum relativistic systems.
Combining last equation and eq.(\ref{cu2.1}) we obtain for the conductivity
tensor $\sigma _{\alpha \alpha }(q,\omega )=\sigma (q,\omega )$%
\begin{equation}
\sigma _{\alpha \beta }(\mathbf{q},\omega )=\frac{\omega \left[ \xi (\mathbf{%
q},\omega )-K_{\alpha \beta }(\mathbf{q},0)\right] }{i\omega
^{2}+iv_{F}^{2}q^{2}}  \label{cu8}
\end{equation}%
The static limit of the homogeneous conductivity can be computed as (see
eq.(40a) of \cite{Janis})%
\begin{equation}
\sigma =\underset{\omega \rightarrow 0}{\lim }\underset{\mathbf{q}%
\rightarrow 0}{\lim }\sigma (\mathbf{q},\omega )=\underset{\omega
\rightarrow 0}{\lim }\frac{1}{i\omega }\xi (0,\omega )-\frac{1}{i\omega }%
\underset{\mathbf{q}\rightarrow 0}{\lim }\underset{\omega \rightarrow 0}{%
\lim }K_{\alpha \beta }(\mathbf{q},\omega )=e^{2}D\frac{\partial n}{\partial
\mu }  \label{cu9}
\end{equation}%
that relates the diffusion constant $D$ with the static conductivity, known
as Einstein relation. At zero temperature, $\frac{\partial n}{\partial \mu }%
=n_{F}$, where $n_{F}$ is the density of states at the Fermi energy. In the
other side, replacing eq.(\ref{cu7}) in eq.(\ref{cu2.1}) we obtain for the
density response function%
\begin{equation}
\chi (\mathbf{q},\omega )=\frac{q^{2}\left[ \xi (\mathbf{q},\omega
)-K_{\alpha \alpha }(\mathbf{q},0)\right] }{e^{2}\omega
^{2}+e^{2}v_{F}^{2}q^{2}}  \label{cu10}
\end{equation}%
which is similar to eq.(42) of \cite{Janis}, but in this case, this relation
is exact.

\subsection{Current and density limits}

To compute the two limits $\omega \rightarrow 0$ and $\mathbf{q}\rightarrow
0 $ for the response functions it is only necessary to study the tensor $%
\Gamma _{ijkl}(\mathbf{q},\omega )$ that can be separated as%
\begin{equation}
\Gamma _{ijkl}(\mathbf{q},\omega )=\Gamma _{ijkl}^{RA}(\mathbf{q},\omega
)+\Gamma _{ijkl}^{AA}(\mathbf{q},\omega )+\Gamma _{ijkl}^{RR}(\mathbf{q}%
,\omega )  \label{dc1}
\end{equation}%
In particular, the $\omega \rightarrow 0$ limit reads%
\begin{equation}
\underset{\omega \rightarrow 0}{\lim }\Gamma _{ijkl}(\mathbf{q},\omega
)=\int \frac{dE}{2\pi i}\rho (E)\left[ \Phi _{lijk}^{RR}(E,E,\mathbf{q}%
)-\Phi _{lijk}^{AA}(E,E,\mathbf{q})\right]  \label{dc2}
\end{equation}%
using that $G^{A}=[G^{R}]^{\ast }$we have%
\begin{equation}
\Phi _{lijk}^{RR}(E,E,\mathbf{q})-\Phi _{lijk}^{AA}(E,E,\mathbf{q})=-2i\Im %
\left[ \Phi _{lijk}^{RR}(E,E,\mathbf{q})\right]  \label{dc3}
\end{equation}%
Finally, taking the $\mathbf{q}\rightarrow 0$ limit and using the Ward
identity (see \cite{janis2}) we obtain%
\begin{equation}
\underset{\mathbf{q}\rightarrow 0}{\lim }\underset{\omega \rightarrow 0}{%
\lim }\Gamma _{ijkl}(\mathbf{q},\omega )=\int \frac{dE}{\pi }\rho (E)\int 
\frac{d^{2}\mathbf{k}}{(2\pi )^{2}}\Im \left[ \frac{\partial g_{R}(\mathbf{k}%
,E)}{\partial E}\right] f_{li}(\mathbf{k})f_{jk}(\mathbf{k})  \label{dc4}
\end{equation}%
where%
\begin{equation}
g_{R}(\mathbf{k},E)=\frac{1}{E-\hbar v_{F}k-i\eta -is}  \label{dc5}
\end{equation}%
and%
\begin{equation}
f(\mathbf{k})=\left( 
\begin{array}{cc}
1 & e^{i\varphi _{\mathbf{k}}} \\ 
e^{-i\varphi _{\mathbf{k}}} & 1%
\end{array}%
\right) =\left( 
\begin{array}{cc}
1 & e^{i\lambda } \\ 
e^{-i\lambda } & 1%
\end{array}%
\right)  \label{dc6}
\end{equation}%
where $\lambda $ is the polar angle of the wave vector. Applying the chain
rule in the derivate%
\begin{equation}
\underset{\mathbf{q}\rightarrow 0}{\lim }\underset{\omega \rightarrow 0}{%
\lim }\Gamma _{ijkl}(\mathbf{q},\omega )=-\int \frac{dE}{\pi }\frac{\partial
\rho }{\partial \mu }\int \frac{d^{2}\mathbf{k}}{(2\pi )^{2}}\Im \left[
g_{R}(\mathbf{k},E)\right] f_{li}(\mathbf{k})f_{jk}(\mathbf{k})  \label{dc7}
\end{equation}%
Those integral matrix elements that contains $e^{\pm i\lambda }$ will not
contribute because the angular integration vanish. Multypling last result
with $e^{2}v_{F}^{2}\sigma _{ij}^{\alpha }\sigma _{kl}^{\alpha }$ and taking
the $T\rightarrow 0$ limit we obtain 
\begin{equation}
\underset{\mathbf{q}\rightarrow 0}{\lim }\underset{\omega \rightarrow 0}{%
\lim }\left( K_{\alpha \alpha }^{AA}(\mathbf{q},\omega )+K_{\alpha \alpha
}^{RR}(\mathbf{q},\omega )\right) =-\underset{\mathbf{q}\rightarrow 0}{\lim }%
\underset{\omega \rightarrow 0}{\lim }K_{\alpha \alpha }(\mathbf{q},\omega
)=-e^{2}v_{F}^{2}n_{F}(\eta )  \label{dc8}
\end{equation}%
The longitudinal conductivity depends only in the electron-hole correlation
function as it is expected:%
\begin{equation}
\underset{\mathbf{q}\rightarrow 0}{\lim }\underset{\omega \rightarrow 0}{%
\lim }\sigma (\mathbf{q},\omega )=\underset{\omega \rightarrow 0}{\lim }%
\frac{K_{\alpha \alpha }^{RA}(0,\omega )}{i\omega }  \label{dc8.1}
\end{equation}%
In the other side, taking the $\mathbf{q}\rightarrow 0$ limit and using the
Ward identity we obtain 
\begin{gather}
\underset{\mathbf{q}\rightarrow 0}{\lim }\Gamma _{ijkl}(\mathbf{q},\omega
)=\int \frac{dE}{2\pi i}\int \frac{d^{2}\mathbf{k}}{(2\pi )^{2}}[\rho
(E+\hbar \omega )\left( g_{A}(\mathbf{k},E+\hbar \omega )-g_{R}(\mathbf{k}%
,E+\hbar \omega )\right) -  \label{dc8.2} \\
\rho (E)\left( g_{A}(\mathbf{k},E)-g_{R}(\mathbf{k},E)\right) ]f_{li}(%
\mathbf{k})f_{jk}(\mathbf{k})=0  \notag
\end{gather}%
Because both contributions gives the density of states at the Fermi level
when the tensor $f_{li}(\mathbf{k})f_{jk}(\mathbf{k})$ is contracted with $%
\Lambda _{ijkl}^{\alpha \beta }$. Last equation and the result of eq.(\ref%
{dc8}) implies that the tensor $\Gamma _{ijkl}(\mathbf{q},\omega )$ is not
analytical in the $\mathbf{q}\rightarrow 0$ and $\omega \rightarrow 0$ limit
as it occurs in conventional systems.

\subsection{Boltzmann limit and minimum conductivity}

The Boltzmann limit can be introduced by making the following approximation
(see \cite{Rammer})%
\begin{gather}
\left\langle G_{li}^{R(A)}(\mathbf{k}+\mathbf{q},\mathbf{k}^{\prime }+%
\mathbf{q},E)G_{jk}^{R(A)}(\mathbf{k}^{\prime }-\mathbf{q},\mathbf{k}-%
\mathbf{q},E^{\prime })\right\rangle \sim  \label{au1} \\
\left\langle G_{li}^{R(A)}(\mathbf{k}+\mathbf{q},\mathbf{k}^{\prime }+%
\mathbf{q},E)\right\rangle \left\langle G_{jk}^{R(A)}(\mathbf{k}^{\prime }-%
\mathbf{q},\mathbf{k}-\mathbf{q},E^{\prime })\right\rangle  \notag \\
=\delta _{k,k^{\prime }}G_{li}^{R(A)}(\mathbf{k}+\mathbf{q},E)G_{jk}^{R(A)}(%
\mathbf{k}-\mathbf{q},E^{\prime })  \notag
\end{gather}%
where $G_{li}^{R(A)}(\mathbf{k}+\mathbf{q},E)$ is the impurity averaged
Green function computed in Section I. Because in the $\omega \rightarrow 0$
limit, the conductivity will depends on the electron-hole correlation
function $\Phi _{lijk}^{RA}$, we will compute $\Gamma _{ijkl}^{RA}(\mathbf{q}%
,\omega )$. Introducing a shift $E\rightarrow E-\frac{\hbar \omega }{2}$ we
have%
\begin{equation}
\Gamma _{ijkl}^{RA}(\mathbf{q},\omega ,\eta )=-i\int \frac{dE}{2\pi }\left[
\rho (E+\frac{\hbar \omega }{2})-\rho (E-\frac{\hbar \omega }{2})\right]
\int \frac{d^{2}\mathbf{k}}{(2\pi )^{2}}G_{li}^{R}(\mathbf{k-q},E+\frac{%
\hbar \omega }{2})G_{jk}^{A}(\mathbf{k}-\mathbf{q},E-\frac{\hbar \omega }{2})
\label{au2}
\end{equation}%
Because we have to compute the trace $\Lambda _{ijkl}^{\alpha \alpha }\Gamma
_{ijkl}(\mathbf{q},\omega ,\eta )$, the only tensor elements that are not
zero reads%
\begin{gather}
\xi (\mathbf{q},\omega )=e^{2}v_{F}^{2}(\Lambda _{1212}^{\alpha \alpha
}\Gamma _{1212}^{RA}+\Lambda _{1221}^{\alpha \alpha }\Gamma
_{1221}^{RA}+\Lambda _{2112}^{\alpha \alpha }\Gamma _{2112}^{RA}+\Lambda
_{2121}^{\alpha \alpha }\Gamma _{2121}^{RA}+  \label{au3} \\
\Lambda _{1111}^{\alpha \alpha }\Gamma _{1111}^{RA}+\Lambda _{1122}^{\alpha
\alpha }\Gamma _{1122}^{RA}+\Lambda _{2211}^{\alpha \alpha }\Gamma
_{2211}^{RA}+\Lambda _{2222}^{\alpha \alpha }\Gamma _{2222}^{RA})  \notag
\end{gather}%
where%
\begin{equation}
\Lambda _{1212}^{\alpha \alpha }=\sigma _{12}^{\alpha }\sigma _{12}^{\alpha
}=e^{2}v_{F}^{2}e^{2i\phi }\text{ \ \ \ \ \ \ \ \ \ \ \ \ \ }\Lambda
_{2121}^{\alpha \alpha }=\sigma _{21}^{\alpha }\sigma _{21}^{\alpha
}=e^{2}v_{F}^{2}e^{-2i\phi }  \label{au4}
\end{equation}%
\begin{equation}
\Lambda _{1221}^{\alpha \alpha }=\Lambda _{2112}^{\alpha \alpha }=\Lambda
_{1111}^{\alpha \alpha }=\Lambda _{1122}^{\alpha \alpha }=\Lambda
_{2211}^{\alpha \alpha }=\Lambda _{2222}^{\alpha \alpha }=e^{2}v_{F}^{2}
\label{au5}
\end{equation}%
In turn%
\begin{equation}
\Gamma _{1221}^{RA}=\Gamma _{2112}^{RA}=\Gamma _{1111}^{RA}=\Gamma
_{2222}^{RA}=g^{RA}(k,q,E,\omega )  \label{au7}
\end{equation}%
and%
\begin{eqnarray}
\Gamma _{2121}^{RA} &=&g^{RA}(\mathbf{k},\mathbf{q},E,\omega )e^{i(\varphi _{%
\mathbf{k}+\mathbf{q}}+\varphi _{\mathbf{k}-\mathbf{q}})}\text{ \ \ \ \ \ \
\ \ \ \ \ }\Gamma _{1122}^{RA}=g^{RA}(\mathbf{k},\mathbf{q},E,\omega
)e^{i(\varphi _{\mathbf{k}-\mathbf{q}}-\varphi _{\mathbf{k}+\mathbf{q}})}
\label{au8} \\
\Gamma _{2211}^{RA} &=&g^{RA}(\mathbf{k},\mathbf{q},E,\omega )e^{i(\varphi _{%
\mathbf{k}+\mathbf{q}}-\varphi _{\mathbf{k}-\mathbf{q}})}\text{ \ \ \ \ \ \
\ \ \ \ \ }\Gamma _{1212}^{RA}=g^{RA}(\mathbf{k},\mathbf{q},E,\omega
)e^{-i(\varphi _{\mathbf{k}+\mathbf{q}}+\varphi _{\mathbf{k}-\mathbf{q}})} 
\notag
\end{eqnarray}%
where%
\begin{gather}
g^{RA}(\mathbf{k},\mathbf{q},E,\omega )=g_{R}(\mathbf{k}+\mathbf{q},E+\frac{%
\hbar \omega }{2})g_{A}(\mathbf{k}-\mathbf{q},E-\frac{\hbar \omega }{2})=
\label{au12} \\
\frac{1}{\left( E+\frac{\hbar \omega }{2}-\hbar v_{F}\left\vert \mathbf{k}+%
\mathbf{q}\right\vert -i\eta -is\right) \left( E-\frac{\hbar \omega }{2}%
-\hbar v_{F}\left\vert \mathbf{k}-\mathbf{q}\right\vert -i\eta +is\right) } 
\notag
\end{gather}%
For $\mathbf{q}=0$%
\begin{gather}
\xi ^{RA}(0,\omega )=\Lambda _{ijkl}^{\alpha \alpha }\Gamma
_{ijkl}^{RA}(0,\omega )=-e^{2}v_{F}^{2}\int \frac{dE}{2\pi }\left[ \rho (E+%
\frac{\hbar \omega }{2})-\rho (E-\frac{\hbar \omega }{2})\right]
\label{au13} \\
\int \frac{d^{2}\mathbf{k}}{(2\pi )^{2}}\frac{e^{2i\phi }e^{-2i\varphi _{%
\mathbf{k}}}+e^{-2i\phi }e^{2i\varphi _{\mathbf{k}}}+2}{\left( E+\frac{\hbar
\omega }{2}-\hbar v_{F}k-i\eta -is\right) \left( E-\frac{\hbar \omega }{2}%
-\hbar v_{F}k-i\eta +is\right) }  \notag
\end{gather}%
writing%
\begin{equation}
e^{\pm 2i\varphi \mathbf{_{\mathbf{k}}}}=\cos (2arctg(\frac{k_{y}}{k_{x}}%
))+i\sin (2arctg(\frac{k_{y}}{k_{x}}))=\frac{1}{k^{2}}\left(
k_{x}^{2}-k_{y}^{2}\pm 2ik_{x}k_{y}\right)  \label{au14}
\end{equation}%
using that $k_{x}=k\cos \lambda $ and $k_{y}=k\sin \lambda $ and computing
the angular integration, we obtain%
\begin{gather}
\xi ^{RA}(0,\omega )=\Lambda _{ijkl}^{\alpha \alpha }\Gamma
_{ijkl}^{RA}(0,\omega )=-2e^{2}v_{F}^{2}\int \frac{dE}{2\pi }\left[ \rho (E+%
\frac{\hbar \omega }{2})-\rho (E-\frac{\hbar \omega }{2})\right]
\label{au15} \\
\int \frac{d^{2}\mathbf{k}}{(2\pi )^{2}}\frac{1}{\left( E+\frac{\hbar \omega 
}{2}-\hbar v_{F}k-i\eta -is\right) \left( E-\frac{\hbar \omega }{2}-\hbar
v_{F}k-i\eta +is\right) }  \notag
\end{gather}%
The small parameter $is$ can be disregard because the self-energy $i\eta $
moves the poles of $\xi ^{RA}(0,\omega )$ away from the real line. Using a
simplified version of the Ward identity, we can compute the integral in $k$
as follows 
\begin{gather}
\frac{1}{2\pi }\int_{0}^{1/a}\frac{kdk}{(E+\frac{\hbar \omega }{2}%
-v_{f}\hbar k-i\eta )(E-\frac{\hbar \omega }{2}-v_{f}\hbar k-i\eta )}=
\label{au16} \\
\frac{1}{2\pi \hbar \omega }\int_{0}^{1/a}dk\left( \frac{k}{E+\frac{\hbar
\omega }{2}-v_{f}\hbar k-i\eta }-\frac{k}{E-\frac{\hbar \omega }{2}%
-v_{f}\hbar k-i\eta }\right)  \notag
\end{gather}%
A special feature about graphene is the no disorder limit $\eta \rightarrow
0 $. In this case, the density of states at the Fermi level is zero $%
n(E_{F})=0 $ which implies that there is no charge carriers. Nevertheless, a
minimal conductivity value can be found as follows:\ last equation can be
separated in a real and imaginary part, but the principal part will not
contribute to the conducticity because it vanishes since $\frac{k(E+\frac{%
\hbar \omega }{2}-v_{f}\hbar k)}{(E+\frac{\hbar \omega }{2}-v_{f}\hbar
k)^{2}+\eta ^{2}}$ is an odd function of $E\pm \frac{\hbar \omega }{2}%
-v_{f}\hbar k$, the width $\eta $ is small and $k$ is a slow varying
function from~$0$ to $1/a$, then%
\begin{equation}
\underset{\eta \rightarrow 0}{\lim }\frac{i}{2\pi \hbar \omega }%
\int_{0}^{1/a}dk\left[ \frac{k\eta }{(E+\frac{\hbar \omega }{2}-v_{f}\hbar
k)^{2}+\eta ^{2}}-\frac{k\eta }{(E-\frac{\hbar \omega }{2}-v_{f}\hbar
k)^{2}+\eta ^{2}}\right] =-\frac{i}{2v_{F}^{2}\hbar ^{2}}  \label{au17}
\end{equation}%
Using last result in eq.(\ref{au15}) and taking into account that%
\begin{equation}
\rho (E+\frac{\hbar \omega }{2})-\rho (E-\frac{\hbar \omega }{2})=-\frac{%
\sinh (\frac{\beta \hbar \omega }{2})}{\cosh (\frac{\beta \hbar \omega }{2}%
)+\cosh (\beta E)}  \label{au19}
\end{equation}%
which behaves at low temperatures as $\rho (E+\frac{\hbar \omega }{2})-\rho
(E-\frac{\hbar \omega }{2})\sim 1\,$between $-\frac{\hbar \omega }{2}$ and $%
\frac{\hbar \omega }{2}$ and zero in the remaining energy values, then%
\begin{equation}
\int \frac{dE}{2\pi }\left[ \rho (E+\frac{\hbar \omega }{2})-\rho (E-\frac{%
\hbar \omega }{2})\right] b(E)=\int_{-\frac{\hbar \omega }{2}}^{\frac{\hbar
\omega }{2}}b(E)dE  \label{au20}
\end{equation}%
where $b(E)$ is any function. Eq.(\ref{au15}) finally reads%
\begin{equation}
\xi ^{RA}(0,\omega )=\Lambda _{ijkl}^{\alpha \alpha }\Gamma
_{ijkl}^{RA}(0,\omega ,\eta )=\frac{ie^{2}\hbar \omega }{2\pi \hbar ^{2}}
\label{au21}
\end{equation}%
by applying eq.(\ref{cu9})%
\begin{equation}
\sigma _{0}=\frac{1}{i\omega }\frac{ie^{2}\hbar \omega }{2\pi \hbar ^{2}}=%
\frac{e^{2}}{2\pi \hbar }  \label{au22}
\end{equation}%
Altough there is no disorder ($\eta \rightarrow 0~$limit) and in consequence
no density of states at the Fermi energy, is unusual to obtain a minimal
conductivity. This result is agreement with the result found in (\cite{Peres}%
, eq.(2.53)), but in disagreement with other results (see \cite{Ziegler}).%
\footnote{%
The conductivity of eq.(\ref{au22}) must be multiplied by the degeneracy
given by spin and valley $K$ and $K^{\prime }$. Then, the value would be $%
4\sigma _{0}$.} As we point before, we are using the Born approximation to
treat impurity effects in graphene, which is valid only in the weak
scattering regime. This impose conditions on the possible value of the the
impurity potential $V_{0}$, in particular, it should be less than the
bandwidth because we are in the linear dispersion regime. In turn, this
approximation ommit scatterings on pairs and larger groups of impurities,
then it is expected to remain valid provided cluster effects are
insignificant. In the other side, when impurity concentration is gradually
increased, individual impurity states begin to overlap and the contribution
from these states to the self--energy is becoming more pronounced in the
vicinity of the impurity state energy and a spectrum rearrangement appears
for a critical concentration (see \cite{yuriv}). This impose several
restrictions to the possible values for the the concentration of impurities
and the potential $V_{0}$ value (see \cite{yuriv2} and \cite{yuriv3}), which
in turn impose several restrictions to the approximation used in this work,
because it cannot be applied in a close vicinity of the Dirac point in the
spectrum due to the increase in cluster scattering. Nevertheless, in \cite%
{loktev1} and \cite{loktev2}, a $E_{F}\rightarrow 0$ limit is taken on the
average Green function and by using the Ioffe-Regel criterion (see \cite%
{ioffe}), one of the solutions of this limit implies that the
self-consistent method is not applicable near the nodal point, which is
equivalent to the conditions found in \cite{yuriv2}, but another low energy
asymptotics solution exists, which impose more suitable conditions for the
applicability of the Born approximation (see eq.(9) of \cite{loktev2}). This
point deserves more attention, because the low energy limit in the graphene
Green function and correlation functions raise up a non-analytical behavior
which produces different results (see \cite{Ziegler}). Another important
point is to compute minimum conductivity by taking into account the Velick%
\'{y}-Ward identity, which introduce a two-particle irreducible vertex
consistent with the coherent-potential approximation for the self-energy
(see \cite{veli}, \cite{baym} and \cite{baym2}). In particular, a Cooper
pole could be computed in the two-particle irreducible vertex due to
backscattering, which will dominate the low-energy behavior of the
conductivity and this could give some insight for the minimum conductivity
puzzle.

\section{Dynamical diffusion}

A dynamical generalization of the diffusion constant from the electron-hole
correlation function cannot be linked to the frequency dependent
conductivity (see eq.(3.18) and eq.(3.19) of \cite{Janis}). For this, is
necessary to obtain a dynamical diffusion from a different procedure. The
relaxation of a non-equilibrium particle density distribution can be studied
through the diffusion equation 
\begin{equation}
\frac{\partial \delta n}{\partial t}-D\nabla ^{2}\delta n=0  \label{diff1}
\end{equation}%
where the Fourier transformed solution reads%
\begin{equation}
\delta n(\mathbf{q},\omega )=\frac{\delta n(t=0,\mathbf{q})}{i\omega -Dq^{2}}
\label{diff2}
\end{equation}%
The induced non-equilibrium density variation that arose as a response to a
weak inhomogeneous electric field, where this perturbation is first slowly
switched on during the time interval $(-\infty ,0)$ and then suddenly turned
off at $t=0$ reads (see \cite{belitz})%
\begin{equation}
\delta n(\mathbf{q},t)=eV(\mathbf{q})\theta (t)\int_{-\infty }^{0}dt^{\prime
}e^{\epsilon t^{\prime }}\chi (\mathbf{q},t-t^{\prime })=eV(\mathbf{q})\phi (%
\mathbf{q},t)  \label{diff3}
\end{equation}%
where $V(\mathbf{q})$ is the Fourier transform of the scalar potential and $%
\phi (\mathbf{q},t)$ is the relaxation function. The Fourier transform of
last equation gives a relation between $\delta n(\mathbf{q},\omega )$ and $%
\phi (\mathbf{q},\omega )$, then%
\begin{equation}
\phi (\mathbf{q},\omega )=\frac{\frac{\partial n}{\partial \mu }}{-i\omega
+D(\omega )q^{2}}  \label{diff4}
\end{equation}%
where%
\begin{equation}
\frac{\partial n}{\partial \mu }=\frac{\delta n(t=0,\mathbf{q})}{eV(\mathbf{q%
})}  \label{diff5}
\end{equation}%
From eq.(\ref{diff4}) we can obtain the dynamical diffusion 
\begin{equation}
2\frac{\partial n}{\partial \mu }D(\omega )=\omega ^{2}\frac{\partial
^{2}\phi }{\partial \mathbf{q}^{2}}\mid _{\mathbf{q}=0}  \label{diff6}
\end{equation}%
In turn, from eq.(\ref{diff3}) we obtain a relation between the relaxation
function $\phi (\omega ,q)$ and the response function $\chi (\omega ,q)$ 
\begin{equation}
i\omega \phi (\mathbf{q},\omega )=\chi (\mathbf{q},\omega )-\chi (\mathbf{q}%
,0)  \label{diff7}
\end{equation}%
Using eq.(\ref{cu10}), we can obtain the dynamical diffusion in terms of $%
\xi (\mathbf{q},\omega )$ without taking the $\eta \rightarrow 0$ limit%
\begin{equation}
2\frac{\partial n}{\partial \mu }D(\omega ,\eta )=\frac{2}{ie^{2}\omega }%
(\xi (0,\omega )-\underset{\mathbf{q}\rightarrow 0}{\lim }\underset{\omega
\rightarrow 0}{\lim }K_{\alpha \alpha }(\mathbf{q},\omega ))+\frac{i\omega }{%
e^{2}v_{F}^{2}}(\frac{\partial ^{2}K_{\alpha \alpha }(\mathbf{q},0)}{%
\partial \mathbf{q}^{2}}\mid _{\mathbf{q}=0}-\frac{\partial ^{2}\xi (\mathbf{%
q},0)}{\partial \mathbf{q}^{2}}\mid _{\mathbf{q}=0})  \label{diff8}
\end{equation}%
The dynamical diffusion will contain two contributions at order $O(\omega )$%
. The first one contains the diffusion pole $1/\omega $ of the relaxation
function and will not depends on disorder. The second term will be
proportional to $\omega $ and the factor will be a $\eta $ dependent
function. From last section, we found that $\xi (0,\omega )=0$ and that $%
\underset{\mathbf{q}\rightarrow 0}{\lim }\underset{\omega \rightarrow 0}{%
\lim }K_{\alpha \alpha }(\mathbf{q},\omega )=e^{2}v_{F}^{2}n_{F}$, then%
\begin{equation}
2\frac{\partial n}{\partial \mu }D(\omega )=-\frac{2v_{F}^{2}n_{F}}{i\omega }%
-i\omega \frac{\partial ^{2}\Gamma _{iill}(\mathbf{q},0)}{\partial \mathbf{q}%
^{2}}\mid _{\mathbf{q}=0}  \label{diff8.1}
\end{equation}%
where we have used eq.(\ref{cu2.1}) to eq.(\ref{cu5}). The $\eta $ dependent
factor will depends on the electron-hole correlation function, but in this
case, we have take into account the $\mathbf{q}$ dependence. Using eq.(\ref%
{dc4}), last term of the r.h.s. of eq.(\ref{dc2}) can be written as%
\begin{equation}
\frac{\partial ^{2}\Gamma _{iill}(\mathbf{q},0)}{\partial \mathbf{q}^{2}}%
\mid _{\mathbf{q}=0}=-\int \frac{dE}{\pi }\rho (E)\frac{\partial ^{2}\Im %
\left[ \Phi _{liil}^{RR}(E,E,\mathbf{q})\right] }{\partial \mathbf{q}^{2}}%
\mid _{\mathbf{q}=0}  \label{diff8.2}
\end{equation}%
Writing 
\begin{equation}
\frac{\partial ^{2}\Im \left[ \Phi _{liil}^{RR}(E,E,\mathbf{q})\right] }{%
\partial \mathbf{q}^{2}}\mid _{\mathbf{q}=0}=\frac{dS}{dE}  \label{diff8.3}
\end{equation}%
Eq.(\ref{diff8.2}) can be written as%
\begin{equation}
\frac{\partial ^{2}\Gamma _{iill}(\mathbf{q},0)}{\partial \mathbf{q}^{2}}%
\mid _{\mathbf{q}=0}=-\int \frac{dE}{\pi }\rho (E)\frac{\partial ^{2}\Im %
\left[ \Phi _{liil}^{RR}(E,E,\mathbf{q})\right] }{\partial \mathbf{q}^{2}}%
\mid _{\mathbf{q}=0}=\frac{1}{\pi }S(-\infty )+\frac{S(0)}{\pi }
\label{diff8.4}
\end{equation}%
where we have integrate by parts and used that $\frac{\partial \rho }{%
\partial E}=\delta (E)$ in the $T\rightarrow 0$ limit. \ Taking into account
eq.(\ref{con20.6.1.7}) in the Boltzmann limit, the electron-electron
correlation function $\Phi _{liil}^{RR}(E,E,\mathbf{q})$ can be written as%
\begin{gather}
\Phi _{liil}^{RR}(E,E,\mathbf{q})=  \label{diff8.5} \\
\int_{0}^{1/a}\int_{0}^{2\pi }\frac{dkkd\lambda }{(2\pi )^{2}}\frac{1+\frac{%
k^{2}-q^{2}}{\sqrt{(k^{2}-2qk\cos \lambda +q^{2})(k^{2}+2qk\cos \lambda
+q^{2})}}}{(E-\hbar v_{F}\sqrt{k^{2}+2kq\cos \lambda +q^{2}}-i\eta )(E-\hbar
v_{F}\sqrt{k^{2}-2kq\cos \lambda +q^{2}}-i\eta )}  \notag
\end{gather}%
where we have put the $\mathbf{q}$ direction in the same direction as $k_{x}$%
, then $\mathbf{q\cdot k}=qk\cos \lambda $ where $\lambda $ is the polar
angle of $\mathbf{k}$. In appendix C we have computed $\frac{\partial
^{2}\Im \left[ \Phi _{liil}^{RR}(E,E,q)\right] }{\partial \mathbf{q}^{2}}%
\mid _{\mathbf{q}=0}$, where the result reads%
\begin{equation}
\frac{\partial ^{2}\left[ \Phi _{liil}^{RR}(E,E,\mathbf{q})\right] }{%
\partial \mathbf{q}^{2}}\mid _{\mathbf{q}=0}=\frac{1}{\pi }\int_{0}^{1/a}dk%
\left[ -\hbar v_{F}g_{R}^{3}(k,E)+k\hbar ^{2}v_{F}^{2}g_{R}^{4}(k,E)-\frac{%
g_{R}^{2}(k,E)}{k}\right]  \label{diff8.5.1}
\end{equation}%
Last integral will contain give a divergent result in the limit $%
k\rightarrow 0$, which is an infrared divergence due to the masless behavior
of electrons. To isolate the divergence, we can expand the integral in
powers of $k$ before introducing the integral limits%
\begin{gather}
\frac{\partial ^{2}\left[ \Phi _{liil}^{RR}(E,E,\mathbf{q})\right] }{%
\partial \mathbf{q}^{2}}\mid _{\mathbf{q}=0}=-\frac{10+6iarctg(\frac{\eta }{E%
})}{6\pi (E-i\eta )^{2}}+\frac{1}{2\pi (E-i\eta )^{2}}\ln (\frac{E^{2}+\eta
^{2}}{\hbar ^{2}v_{F}^{2}k^{2}})  \label{diff8.6} \\
+\frac{1}{\pi }\underset{j=1}{\overset{+\infty }{\sum }}b_{j}\frac{(\hbar
v_{F}k)^{j}}{(E-i\eta )^{j+2}}  \notag
\end{gather}%
where 
\begin{equation}
b_{j}=\frac{j^{3}-3j^{2}-10j-6}{6j}  \label{diff8.6.1}
\end{equation}%
Introducing a lower cutoff $\Lambda $, integral of eq.(\ref{diff8.5.1}) reads%
\begin{equation}
\frac{\partial ^{2}\left[ \Phi _{liil}^{RR}(E,E,\mathbf{q})\right] }{%
\partial \mathbf{q}^{2}}\mid _{\mathbf{q}=0}=\frac{\ln (a\Lambda )}{\pi
(E-i\eta )^{2}}+\frac{1}{\pi }\underset{j=1}{\overset{+\infty }{\sum }}b_{j}%
\frac{(\hbar v_{F})^{j}}{(E-i\eta )^{j+2}}(\frac{1}{a^{j}}-\Lambda ^{j})
\label{diff8.7}
\end{equation}%
Taking the imaginary part of eq.(\ref{diff8.7})%
\begin{gather}
\Im \frac{\partial ^{2}\left[ \Phi _{liil}^{RR}(E,E,\mathbf{q})\right] }{%
\partial \mathbf{q}^{2}}\mid _{\mathbf{q}=0}=\frac{2E\eta \ln (a\Lambda )}{%
\pi (E^{2}+\eta ^{2})^{2}}  \label{diff8.9} \\
+\frac{1}{\pi }\underset{j=1}{\overset{+\infty }{\sum }}b_{j}(\hbar
v_{F})^{j}(E^{2}+\eta ^{2})^{-\frac{1}{2}(j+2)}\sin ((j+2)arctg(\frac{\eta }{%
E}))(\frac{1}{a^{j}}-\Lambda ^{j})  \notag
\end{gather}%
Integrating in $E$ and taking the two limits of eq.(\ref{diff8.4})%
\begin{equation}
\frac{\partial ^{2}\Gamma _{iill}(\mathbf{q},0)}{\partial \mathbf{q}^{2}}%
\mid _{\mathbf{q}=0}=-\frac{\ln (a\Lambda )}{\pi \eta }-\frac{1}{\pi }%
\underset{j=1}{\overset{+\infty }{\sum }}b_{j}\frac{\sin (1+j)\frac{\pi }{2})%
}{j+1}\frac{(\hbar v_{F})^{j}}{\eta ^{j+1}}(\frac{1}{a^{j}}-\Lambda ^{j})
\label{diff8.10}
\end{equation}%
Last equation depends on the lower cutoff $\Lambda $, which is not desired.
A correct procedure can be applied by assuming that the Fermi velocity $%
v_{F} $ will change with $\Lambda $.\footnote{%
The Fermi velocity is one of the parameters of the Hamiltonian.} A
renormalization group equation can be obtained by assuming that the
dynamical diffusion do not depends on $\Lambda $. Then%
\begin{gather}
\frac{4n_{F}v_{F}}{i\omega }\frac{dv_{F}}{d\Lambda }+i\omega \frac{\partial 
}{\partial v_{F}}\left( \frac{\partial ^{2}\Gamma _{iill}(\mathbf{q},0,\eta
,\Lambda ,v_{F}(\Lambda ))}{\partial \mathbf{q}^{2}}\mid _{\mathbf{q}%
=0}\right) \frac{dv_{F}}{d\Lambda }+  \label{diff8.11} \\
i\omega \frac{\partial }{\partial \Lambda }\left( \frac{\partial ^{2}\Gamma
_{iill}(\mathbf{q},0,\eta ,\Lambda ,v_{F}(\Lambda ))}{\partial \mathbf{q}^{2}%
}\mid _{\mathbf{q}=0}\right) =0  \notag
\end{gather}%
Eq.(\ref{diff8.10}) is suitable to compute differents orders of $\hbar $ to
the renormalization group equation for $v_{F}$. At order $\hbar ^{0}$ we
obtain

\begin{figure}[tbp]
\centering
\includegraphics[width=105mm,height=70mm]{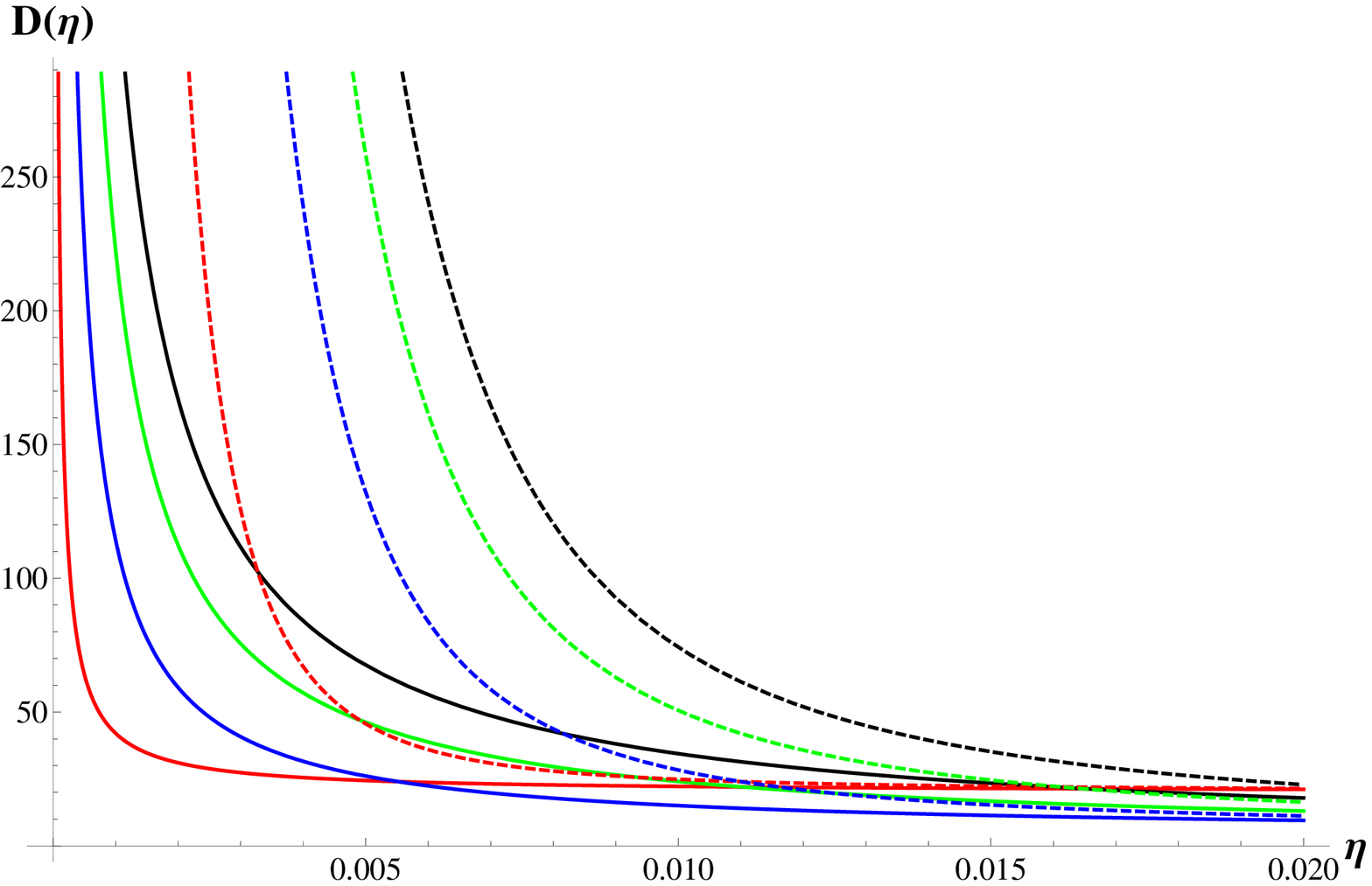}
\caption{Dynamical diffusion as a function of disorder for different values
of $\protect\omega $ in arbitrary units. From red to black solid lines, $%
\protect\omega $ increase. Dashed lines for dynamical diffusion at order $%
O(\hbar ^{2})$. }
\label{dynamical1}
\end{figure}

\begin{equation}
-\frac{4v_{F}n_{F}}{i\omega }\frac{dv_{F}}{d\Lambda }+\frac{i\omega }{\pi
\eta }\ln (a\Lambda )=0  \label{diff8.12}
\end{equation}%
and the solution reads\footnote{%
In eq.(\ref{diff8.13}) $\Lambda _{0}=1$ as a low limit of the cutoff has
been used.}%
\begin{equation}
v_{F}(\Lambda )=\sqrt{v_{F_{0}}^{2}-\frac{\omega ^{2}}{2\pi \eta n_{F}}\ln
(\Lambda )}  \label{diff8.13}
\end{equation}%
because we are in the approximation $\omega \rightarrow 0$, last equation
reads%
\begin{equation}
v_{F}(\Lambda )=v_{F_{0}}-\frac{\omega ^{2}}{4\pi n_{F}\eta v_{F}^{(0)}}\ln
(\Lambda )  \label{diff8.13.1}
\end{equation}%
which is similar to the results found in (\cite{atta}, \cite{song}) and
shows a singular behavior with the impurity factor $\eta $ that is similar
to the singular behavior of the Fermi velocity with impurities found in \cite%
{yuriv4}.\footnote{%
If we introduce a upper cutoff $k_{\infty }=\frac{1}{\Lambda }$, the result
of eq.(\ref{diff8.13}) follows the same behavior as other results.} Using
eq.(\ref{diff8.13}), the dynamical diffusion at order $\hbar ^{0}$ reads%
\begin{equation}
D(\omega )=\frac{iv_{F_{0}}^{2}}{\omega }+\frac{i\omega }{2\pi \eta n_{F}}%
\ln (a)  \label{diff8.14}
\end{equation}%
\begin{figure}[tbp]
\centering
\includegraphics[width=105mm,height=70mm]{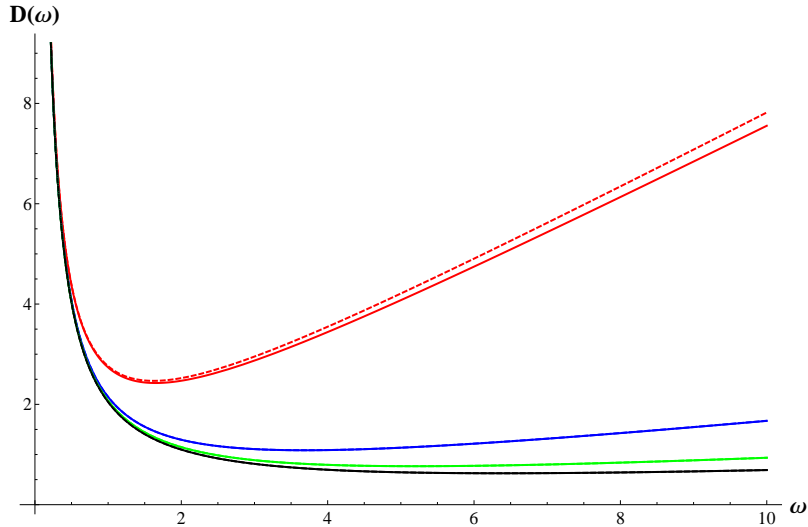}
\caption{Dynamical diffusion as a function of $\protect\omega $ for
different values of $\protect\eta $ in arbitrary units. From red to black
solid lines, $\protect\eta $ increase. Dashed lines for dynamical diffusion
at order $O(\hbar ^{2})$. }
\label{dynamical2}
\end{figure}

From last equation, there is no real value for $\omega $ where $D(\omega )=0$%
, which is expected because suppression of diffusion can be achieved by
taking into account maximally crossed diagrams in the perturbation
expansion. Nevertheless, we can plot $D(\omega ,\eta )$ as a function of $%
\omega $ for different values of $\eta $ and $D(\omega ,\eta )$ as a
function of $\eta $ for different values of $\omega $. Both figures show the
diffusion pole at $\omega =0$ \ (see \ref{dynamical1} and \ref{dynamical2}).
Dynamical diffusion tends to $\frac{v_{F_{0}}^{2}}{2i\omega }$ for $\eta
\rightarrow \infty $. In turn, dynamical diffusion shows a minimum which
corresponds to the following frequency%
\begin{equation}
\omega =\sqrt{\frac{2\pi v_{F_{0}}^{2}\eta n_{F}}{\ln (a)}}  \label{diff8.15}
\end{equation}%
which is proportional to the impurity potential $V_{0}$. This implies that
at low resonance frequencies, a decrease in the diffusion can be expected.
The full renormalization group equation can be computed in the no-disorder
limit $\eta \rightarrow 0$, which gives the following differential equation
at order $O(\hbar ^{j})$:%
\begin{equation}
\frac{dv_{F}}{v_{F}}=\frac{\Lambda ^{j-1}d\Lambda }{(\frac{1}{a^{j}}-\Lambda
^{j})}  \label{diff8.16}
\end{equation}%
where the solution reads%
\begin{equation}
v_{F}(\Lambda )=v_{F_{0}}(\frac{1-a^{-j}}{\Lambda ^{j}-a^{-j}})^{1/j}
\label{diff8.17}
\end{equation}%
In the cut off limit $\Lambda \rightarrow 0$, the Fermi velocity change as%
\begin{equation}
\frac{v_{F}}{v_{F_{0}}}=a(a^{-j}-1)^{1/j}  \label{diff8.18}
\end{equation}%
which decreases with increasing order of $\hbar $. In turn, Fermi velocity
at higher orders of $\hbar $ and in the no disorder limit do not depends on
the frequency of the external perturbation. High impurity concentrations in
graphene can lead to a diffusion suppresion which would leave without effect
the high performance of the material sample as a gas sensor. Weak
localization of electrons in doped graphene implies to take into account
higher orders in the diagrammatic perturbation expansion of the current
response function. Some theoretical computations has been done (see \cite%
{wl0}, \cite{wl1}). For conventional impurities, the correction becomes
positive and it leads to the fact that anti-localization is realized, which
would enhance the gas sensor peformance. In contrast, negative corrections
for short-range impurities are expected from symmetry consideration. This
suggest that the high sensitivity of graphene to detect individual dopants
is highly dependent on the quantum corrections to the conductivity. Finally,
taking into account the first quantum correction to the renormalization
group equation for $v_{F}$ we obtain%
\begin{equation}
v_{F}dv_{F}=\frac{5\omega \hbar ^{2}v_{F}^{2}}{3\pi \eta ^{3}}\frac{d\Lambda 
}{L(\Lambda ,\omega ,\eta ,a)}-\frac{1}{\pi \eta }\frac{d\Lambda }{\Lambda
^{2}L(\Lambda ,\omega ,\eta ,a)}  \label{diff9}
\end{equation}%
where%
\begin{equation}
L(\Lambda ,\omega ,\eta ,a)=\frac{1}{\Lambda }\left( \frac{4n_{F}}{\omega
^{2}}+\frac{5}{3\pi }\frac{\hbar ^{2}}{\eta ^{3}}(\frac{1}{a^{2}}-\Lambda
^{2})\right)  \label{diff10}
\end{equation}%
Solution of eq.(\ref{diff10}) in a integral form reads%
\begin{equation}
v_{F}^{2}(\Lambda )=v_{F_{0}}^{2}-\frac{2}{\pi \eta }\int_{1}^{\Lambda }%
\frac{d\Lambda ^{\prime }}{\Lambda ^{\prime 2}L(\Lambda ^{\prime },\omega
,\eta ,a)}+\frac{10\omega \hbar ^{2}}{3\pi \eta ^{3}}\int_{1}^{\Lambda }%
\frac{v_{F}^{2}(\Lambda ^{\prime })d\Lambda ^{\prime }}{L(\Lambda ^{\prime
},\omega ,\eta ,a)}  \label{diff11}
\end{equation}%
at order $O(\hbar ^{2})$ we need to compute the second term of r.h.s. of
last equation%
\begin{equation}
\int_{1}^{\Lambda }\frac{d\Lambda ^{\prime }}{\Lambda ^{\prime 2}L(\Lambda
^{\prime },\omega ,\eta ,a)}=\frac{3a^{2}\pi \omega ^{2}\eta ^{3}}{10\hbar
^{2}\omega ^{2}+24a^{2}n_{F}\pi \eta ^{3}}\ln \left[ \frac{\left( 5\hbar
^{2}\omega ^{2}(a^{2}-1)-12a^{2}n_{F}\pi \eta ^{3}\right) \Lambda ^{2}}{%
5\hbar ^{2}\omega ^{2}(a^{2}\Lambda ^{2}-1)-12a^{2}n_{F}\pi \eta ^{3}}\right]
\label{diff12}
\end{equation}%
Introducing $v_{F}^{2}(\Lambda )$ in eq.(\ref{diff11}) inside the integral
of the r.h.s. in the same equation and using eq.(\ref{diff12}), the
dynamical diffusion reads at order $O(\hbar ^{2})$%
\begin{equation}
D(\omega )=\frac{v_{F_{0}}^{2}}{\omega }+\frac{i\omega \ln (a)}{2\pi \eta
n_{F}}+\frac{5\omega }{12\pi }\frac{v_{F_{0}}^{2}\hbar ^{2}}{\eta
^{3}n_{F}a^{2}}  \label{diff13}
\end{equation}%
The correction introduced at order $O(\hbar ^{2})$ can be seen in both
figures as dashed lines. In the case of dynamical diffusion in terms of
frequency, the correction is small and only is appreciable for low values of 
$\eta $. In this sense, quantum corrections to the diffusion do not alter
the behavior under local perturbations at linear order in $\omega $.

\section{Conclusion}

In this work a generalization of linear response theory with Kubo formula
has been introduced for linear dispersion relations and spinor wave
functions. A minimal conductivity can be found in the no disorder limit and
the result is in discordance by a factor of $2$ with other theoretical
results, although there is no consensus of the physical reason of such
value. Using the generalization introduced in the first sections, an exact
relation between current and density response functions can be obtained.
Combining this result with the relation obtained with the continuity
equation, an exact functional form of response functions are obtained, where
in particular, a singular behavior appears at $\omega \rightarrow 0$ and $%
q\rightarrow 0$ limit. Finally, dynamical diffusion is computed through the
relaxation function at low order in $\omega $. A regularization is
introduced to avoid infrared divergences, which introduce a renormalization
group equation for the Fermi velocity. Different contributions to this
equation can be analyzed at different order in $\hbar $. Different results
are obtained which are of importance for local pertubations of graphene
sample.

\section{Acknowledgment}

This paper was partially supported by grants of CONICET (Argentina National
Research Council) and Universidad Nacional del Sur (UNS) and by ANPCyT
through PICT 1770, and PIP-CONICET Nos. 114-200901-00272 and
114-200901-00068 research grants, as well as by SGCyT-UNS., E.A.G. and
P.V.J. are members of CONICET. P.B. and J. S.A. are fellow researchers at
this institution.

The authors are extremely grateful to the referee, whose relevant
observations have greatly improved the final version of this paper.

\section{Appendix}

\subsection{Spectral weight}

The Green function for Dirac fermion systems reads%
\begin{equation}
G_{ij}^{R(A)}(\mathbf{r},\mathbf{r}^{\prime },E)=\int \frac{d^{2}\mathbf{k}}{%
(2\pi )^{2}}\frac{\psi _{i}^{(k)}(\mathbf{r})\psi _{j}^{(k)\dag }(\mathbf{r}%
^{\prime })}{E-v_{f}\hbar k(\mp )is}  \label{con21}
\end{equation}%
we can define the spectral weight as%
\begin{equation}
A_{ij}(\mathbf{r},\mathbf{r}^{\prime },E)=i[G_{ij}^{R}(\mathbf{r},\mathbf{r}%
^{\prime },E)-G_{ij}^{A}(\mathbf{r},\mathbf{r}^{\prime },E)]  \label{con22}
\end{equation}%
If we integrate the spectral weight in the volume we obtain the density of
states%
\begin{equation}
\int d^{2}\mathbf{r}A_{ij}(\mathbf{r},\mathbf{r},E)=2\pi \delta _{ij}\int 
\frac{d^{2}\mathbf{k}}{(2\pi )^{2}}\delta (E-v_{f}\hbar k)=2\pi n(E)\delta
_{ij}  \label{con24}
\end{equation}%
where $n(E)$ is the density of states.

\subsection{Electron-electron correlation function}

The electron-electron hole correlation function reads%
\begin{equation}
\Phi _{liil}^{RR}(E,E,\mathbf{q})=\int_{0}^{1/a}\int_{0}^{2\pi }\frac{%
dkkd\lambda }{(2\pi )^{2}}\varrho (\mathbf{k},\mathbf{q},E)  \label{ee1}
\end{equation}%
where $\varrho (\mathbf{k},\mathbf{q},E)=g_{R}(\left\vert \mathbf{k}+\mathbf{%
q}\right\vert ,E)g_{R}(\left\vert \mathbf{k}-\mathbf{q}\right\vert ,E)\alpha
(\mathbf{k},\mathbf{q})$%
\begin{equation}
g_{R}(\left\vert \mathbf{k}\pm \mathbf{q}\right\vert ,E)=\frac{1}{E-\hbar
v_{F}\left\vert \mathbf{k}\pm \mathbf{q}\right\vert -i\eta }  \label{ee2}
\end{equation}%
and%
\begin{equation}
\alpha (\mathbf{k},\mathbf{q})=1+\frac{k^{2}-q^{2}}{\left\vert \mathbf{k}+%
\mathbf{q}\right\vert \left\vert \mathbf{k}-\mathbf{q}\right\vert }
\label{ee3}
\end{equation}%
We can take the $q$ derivate inside the integral in $k$. Taking into account
that%
\begin{equation}
\frac{\partial g_{R}(\left\vert \mathbf{k}\pm \mathbf{q}\right\vert ,E)}{%
\partial \mathbf{q}}=-\frac{\hbar v_{F}\frac{\partial \left\vert \mathbf{k}%
\pm \mathbf{q}\right\vert }{\partial \mathbf{q}}}{(E-\hbar v_{F}\left\vert 
\mathbf{k}\pm \mathbf{q}\right\vert -i\eta )^{2}}=-\hbar v_{F}\frac{\partial
\left\vert \mathbf{k}\pm \mathbf{q}\right\vert }{\partial \mathbf{q}}%
g_{R}^{2}(\left\vert \mathbf{k}\pm \mathbf{q}\right\vert ,E)  \label{ee4}
\end{equation}%
then 
\begin{equation}
\frac{\partial \varrho (\mathbf{k},\mathbf{q},E)}{\partial \mathbf{q}}%
=\varrho (\mathbf{k},\mathbf{q},E)B(\mathbf{k},\mathbf{q},E)  \label{ee5}
\end{equation}%
where%
\begin{equation}
B(\mathbf{k},\mathbf{q},E)=-\hbar v_{F}\frac{\partial \left\vert \mathbf{k}+%
\mathbf{q}\right\vert }{\partial \mathbf{q}}g_{R}(\left\vert \mathbf{k}+%
\mathbf{q}\right\vert ,E)-\hbar v_{F}\frac{\partial \left\vert \mathbf{k}-%
\mathbf{q}\right\vert }{\partial \mathbf{q}}g_{R}(\left\vert \mathbf{k}-%
\mathbf{q}\right\vert ,E)+\frac{\partial \alpha (\mathbf{k},\mathbf{q})}{%
\partial \mathbf{q}}\frac{1}{\alpha (\mathbf{k},\mathbf{q})}  \label{ee6}
\end{equation}%
The second derivate reads%
\begin{equation}
\frac{\partial ^{2}\varrho (\mathbf{k},\mathbf{q},E)}{\partial \mathbf{q}^{2}%
}=\varrho (\mathbf{k},\mathbf{q},E)\left[ B^{2}(\mathbf{k},\mathbf{q},E)+%
\frac{\partial B(\mathbf{k},\mathbf{q},E)}{\partial \mathbf{q}}\right]
\label{ee7}
\end{equation}%
where%
\begin{gather}
\frac{\partial B(\mathbf{k},\mathbf{q},E)}{\partial \mathbf{q}}%
=g_{R}(\left\vert \mathbf{k}+\mathbf{q}\right\vert ,E)\left[ -\hbar v_{F}%
\frac{\partial ^{2}\left\vert \mathbf{k}+\mathbf{q}\right\vert }{\partial 
\mathbf{q}^{2}}+\hbar ^{2}v_{F}^{2}\left( \frac{\partial \left\vert \mathbf{k%
}+\mathbf{q}\right\vert }{\partial \mathbf{q}}\right) ^{2}g_{R}(\left\vert 
\mathbf{k}+\mathbf{q}\right\vert ,E)\right] +  \label{ee8} \\
g_{R}(\left\vert \mathbf{k}-\mathbf{q}\right\vert ,E)\left[ -\hbar v_{F}%
\frac{\partial ^{2}\left\vert \mathbf{k}-\mathbf{q}\right\vert }{\partial 
\mathbf{q}^{2}}+\hbar ^{2}v_{F}^{2}\left( \frac{\partial \left\vert \mathbf{k%
}-\mathbf{q}\right\vert }{\partial \mathbf{q}}\right) ^{2}g_{R}(\left\vert 
\mathbf{k}-\mathbf{q}\right\vert ,E)\right] +  \notag \\
\frac{1}{\alpha (\mathbf{k},\mathbf{q})}\left[ \frac{\partial ^{2}\alpha (%
\mathbf{k},\mathbf{q})}{\partial \mathbf{q}^{2}}-\frac{1}{\alpha (\mathbf{k},%
\mathbf{q})}\left( \frac{\partial \alpha (\mathbf{k},\mathbf{q})}{\partial 
\mathbf{q}}\right) ^{2}\right]  \notag
\end{gather}%
Finally using that 
\begin{equation}
\frac{\partial \left\vert \mathbf{k}\pm \mathbf{q}\right\vert }{\partial 
\mathbf{q}}=\frac{q\pm k\cos \lambda }{\left\vert \mathbf{k}\pm \mathbf{q}%
\right\vert }  \label{ee9}
\end{equation}%
and that the second derivate reads%
\begin{equation}
\frac{\partial ^{2}\left\vert \mathbf{k}\pm \mathbf{q}\right\vert }{\partial 
\mathbf{q}^{2}}=\frac{1}{\left\vert \mathbf{k}\pm \mathbf{q}\right\vert }-%
\frac{(q\pm k\cos \lambda )^{2}}{\left\vert \mathbf{k}\pm \mathbf{q}%
\right\vert ^{3}}  \label{ee10}
\end{equation}%
Putting $\mathbf{q}=0$ in eq.(\ref{ee7}) 
\begin{equation}
\varrho (k,0,E)=2g_{R}(\mathbf{k},E)g_{R}(\mathbf{k},E)  \label{ee11}
\end{equation}%
and using that $\frac{\partial \alpha (k,\mathbf{q})}{\partial \mathbf{q}}%
\mid _{\mathbf{q}=0}=0$,, 
\begin{equation}
B^{2}(k,0,E)=0  \label{ee12}
\end{equation}%
In turn, 
\begin{equation}
\frac{\partial B(\mathbf{k},\mathbf{q},E)}{\partial \mathbf{q}}=2g_{R}(%
\mathbf{k},E)\left[ -\hbar v_{F}\frac{\sin ^{2}\lambda }{k}+\hbar
^{2}v_{F}^{2}\cos ^{2}\lambda g_{R}(k,E)\right] -\frac{2\sin ^{2}\lambda }{%
k^{2}}  \label{ee13}
\end{equation}%
where we have used that%
\begin{equation}
\frac{\partial ^{2}\alpha (\mathbf{k},\mathbf{q})}{\partial \mathbf{q}^{2}}%
\mid _{\mathbf{q}=0}=-\frac{4\sin ^{2}\lambda }{k^{2}}  \label{ee14}
\end{equation}%
Finally the second derivate of $\varrho (\mathbf{k},\mathbf{q},E)$ at $%
\mathbf{q}=0$ reads%
\begin{equation}
\frac{\partial ^{2}\varrho (\mathbf{k},\mathbf{q},E)}{\partial \mathbf{q}^{2}%
}\mid _{\mathbf{q}=0}=4g_{R}^{2}(\mathbf{k},E)\left[ g_{R}(\mathbf{k},E)%
\left[ -\hbar v_{F}\frac{\sin ^{2}\lambda }{k}+\hbar ^{2}v_{F}^{2}\cos
^{2}\lambda g_{R}(\mathbf{k},E)\right] -\frac{\sin ^{2}\lambda }{k^{2}}%
\right]  \label{ee15}
\end{equation}%
which is the desired result which will be used in Section VI.

\end{document}